\documentclass{eptcs}
\providecommand{\event}{FLACOS 2011}
\usepackage{breakurl} 
    \usepackage[dvips]{color}
\usepackage[ruled,vlined]{algorithm2e}        
\usepackage[pdftex]{graphicx}
\usepackage{floatrow}  % Police contenant les caractères français
\usepackage{stmaryrd}
\usepackage{dsfont}
\usepackage{amssymb}
\usepackage{llncsdoc}
\usepackage{enumitem}
\usepackage{algorithmic}
\usepackage[english]{babel}
\usepackage[utf8]{inputenc}
\usepackage{tikz}
\usepackage{multicol}

\usepackage{caption}
\usepackage[labelfont=bf,textfont={sl,bf},lofdepth,lotdepth]{subfig}
\DeclareCaptionListOfFormat{colon}{#2:}
\usepackage{fixltx2e}

\newcommand{\figbox}[1]{%
  \fbox{%
    \vbox to 3.5
in{%
    \vfil
    \hbox to 2.3in{%
      \hfil
      #1%
      \hfil}%
    \vfil}}}

\makeatletter
  \newcommand{\setcaptype}[1]{%
    \listsubcaptions
    \renewcommand{\@captype}{#1}}

\usetikzlibrary{calc,through,backgrounds}
\usetikzlibrary{matrix}
\usetikzlibrary{fit}
\newenvironment{definition}
{\parindent 0cm
\refstepcounter{def} \textbf{Definition \arabic{def} \hspace{0.2cm}}}
{\nopagebreak[4]}
\newcounter{def}

\newenvironment{proposition}
{\parindent 0cm
\refstepcounter{prop} \textbf{Proposition \arabic{prop} \hspace{0.2cm}}}
{\nopagebreak[4]}
\newcounter{prop}

\newenvironment{theorem}
{\parindent 0cm
\refstepcounter{the} \textbf{Theorem \arabic{the} \hspace{0.2cm}}}
{\nopagebreak[4]}

\newcounter{the}

\newenvironment{lemma}
{\parindent 0cm
\refstepcounter{lem} \textbf{Lemma \arabic{lem} \hspace{0.2cm}}}
{\nopagebreak[4]}

\newcounter{lem}

\title{Handling Conflicts in  Depth-First Search for LTL Tableau to Debug Compliance Based Languages{ \thanks{ The research leading to these results has received funding from the European Community'sSeventh Framework Programme FP7/2007-2013 under grant agreement 215483 (S-Cube).}}}

\author{Francois Hantry \qquad\qquad Mohand-Said Hacid
\institute{Universit\'e Claude Bernard Lyon 1\\ LIRIS CNRS UMR 5205\\ Lyon, France}
\email{\quad francois.hantry@liris.cnrs.fr \quad\qquad mohand-said.hacid@liris.cnrs.fr }
}
\def\titlerunning{Handling Conflicts in DFS for LTL tableau}
\def\authorrunning{F. Hantry,M-S. Hacid}
\begin{document}
\maketitle

\maketitle

\begin{abstract}
%\boldmath
 
Providing adequate tools to tackle the problem  of  inconsistent  compliance rules is a critical research topic.
This problem is of paramount importance to achieve automatic support for early declarative design and to support evolution of rules in contract-based or service-based systems. In this paper we investigate the problem of extracting temporal unsatisfiable cores in order to detect the inconsistent part of a specification. We extend  conflict-driven SAT-solver to provide a new conflict-driven depth-first-search solver for temporal logic. We use this solver to compute LTL unsatisfiable cores without re-exploring the history of the solver.

%It enables us to derive  small  inconsistent set of compliance rules. 
\end{abstract}
% IEEEtran.cls defaults to using nonbold math in the Abstract.
% This preserves the distinction between vectors and scalars. However,
% if the conference you are submitting to favors bold math in the abstract,
% then you can use LaTeX's standard command \boldmath at the very start
% of the abstract to achieve this. Many IEEE journals/conferences frown onf
% math in the abstract anyway.

% no keywords

% For peer review papers, you can put extra information on the cover
% page as needed:
% \ifCLASSOPTIONpeerreview
% \begin{center} \bfseries EDICS Category: 3-BBND \end{center}
% \fi
%
% For peerreview papers, this IEEEtran command inserts a page break and
% creates the second title. It will be ignored for other modes.
%\IEEEpeerreviewmaketitle

\section{Introduction}
Providing adequate tools to tackle the problem  of  inconsistent  compliance rules is a critical research topic.
However, few tools fully  analyze conflicts over underpinning logics of a natural language (eg. temporal logic, deontic logic...). Such  early and declarative specifications can  be critical for specifying policies and requirements in agile and distributed environments. Thus, formal languages for compliance requirements  and their analysis have become critical in many computer science domains (eg business process management, service oriented computing, e-commerce). An ongoing research topic is the analysis of a conflicting set of temporal logic compliance rules.
For instance, Table 1.a gives a toy set of compliance rules. It will be used as a running example in the paper.
All those rules except the last one originate from an ongoing supply contract.
Let us assume that the last one (r3.c) originates from another internal requirement from a supplier.
It comes out that this new requirement entails a conflict with rules (r3.a) and rules (r3.b) shown on Table 1.b.
This example shows the importance of automatic detection of conflicting subsets of compliance rules.
This problem is critical for debugging declarative specifications~\cite{GiblinLMPZ05,MontaliPACMS10},  handling conflicting contracts \cite{FenechPS09}, or tackling unrealizable service compositions \cite{MarconiP09}. There exist several formalisms to deal with  time such as LTL, MSO \cite{Emerson}, TLTL , MTL \cite{AlurH91}. These logics underpine many of modern compliance languages and their associated theories and tools are used to address problems related to verification~\cite{GhoseK07,XuLW08,DamaggioDV11}, service composition~\cite{MarconiP09}, graphical design of property patterns~\cite{MontaliPACMS10,GiblinLMPZ05}.\\
We investigate the problem of efficiently extracting temporal logic unsatisfiable cores for debugging compliance rules. Intuitively, an unsatisfiable core is a conflicting subset of rules. We restrict ourselves to LTL for which many results and efficient  model checking methods exist.
However, the problem of  efficiently detecting a small LTL unsatisfiable core is still open ~\cite{Schuppan10} \cite{CimattiRST07}. Conflict driven methods exist for  SAT-solver algorithms. They provide quite efficient extraction of conflicting rules written in propositional logic \cite{ZM03}. SAT-solvers have been extended (e.g.,Unbounded Model Checking (UMC) SAT-solvers \cite{BiereCCFZ99}) to deal with the more expressive LTL.  For the case of satisfiability\footnote{No  model to  check against a LTL formula}, it consists in searching a lasso-shaped model of length $k \leq 2^{\mathcal{O}(|f|)}$ and in reducing to  boolean SAT problems for increasing $k \in [0,2^{\mathcal{O}(|f|)}]$.  One of critical (and basic) points of current boolean SAT-solvers is their ability of pruning `bad' search space.  It is based on a smart use of boolean propagation. Analyzing the propagation enables to handle conflict  while backtracking and enables to avoid revisiting immediately the same conflict. Learning conflict using conflict clause also avoids revisiting the conflict later. This conflict-driven approach leads easily to the extraction of a core.
 %Since UMC is incremental, works (e.g.,\cite{HeljankoJL05},\cite{GanaiGA05}) have tried to reuse learning clause from a boolean SAT problem to the next one. \cite{SheeranSS00}, address safety property and  \cite{McMillan03} propose to approximate the problem by using Craig interpolant. Moreover, these last techniques for safety UMC assume the  translation of a LTL formula to a safety problem, which is in general not efficient for big size formula \cite{Biere06}. Mainly, the main drawback of UMC is the handling of redundant shifted formulas \cite{GanaiGA05} and the difficult computation of an upper bound for $k$ \cite{JussilaB07}.   
\cite{CimattiRST07} proposes to extract unsatisfiable cores from the UMC  method of \cite{HeljankoJL05}. The authors propose also a `Sat Modulo Theory' like framework applied with symbolic global model checking \cite{BurchCMDH92}, but the conflict handling is not introduced inside the symbolic global model checking. \cite{Schuppan10} analyzes a very expanded\footnote{The expansion disregards boolean conflict} tableau of \cite{GerthPVW95} to define unsatisfiable core but again no analysis of conflict is performed.   Thus, on the contrary to boolean SAT-solver and extended UMC, neither global model checking, nor On-The-Fly techniques handle conflict.  Moreover, in the nineties, resolution \cite{Fisher91} for temporal logic has been proposed to tackle  unfair SCC as minimal `temporal conflict' but to the best of our knowledge current boolean SAT-Solvers(e.g.,\cite{HeljankoJL05},\cite{GanaiGA05}, \cite{SheeranSS00},\cite{JussilaB07} ) have not investigated this idea yet, mainly because they are Breadth-First-Search. But, a drawback of resolution is that any conflict is recorded using resolvent, this entails a too large use of  memory space in contrast to On-The-Fly tableau, symbolic model checking and UMC. In this paper, we propose a new  conflict-driven  depth-first-search solver inspired by SAT-based ones, DFS for tableau and resolution for temporal logic. Furthermore, we show how it is possible to extract a small unsatisfiable core.

\paragraph{Overview of the paper}

Section 2  introduces Background.  Section 3 describes sound technical details of section 4. Section 4 shows the Solver. Section 5 is devoted to the correctness, completeness,extraction of unsatisfiable cores. We conclude in Section 6.

\section{Background}

%\subsection{Basic properties of LTL}

\begin{definition} (Syntax of LTL) 

Let $P$ be a non empty finite set of propositional variables, and $p \in P$. $A$ and $B$ two LTL formulas. A temporal logic formula is inductively built by means of the following rules:
\begin{center}
TRUE \textbar FALSE \textbar
 $ p$ \textbar $A \wedge B $ \textbar $A \vee B $ \textbar $\neg A$\textbar $X (A)$\\
  $A U B$ \textbar $A W B$
\end{center}
Furthermore,  $G(A) = (A) W (FALSE)$ and $F(A) = (TRUE) U (A)$.
\end{definition}

\begin{definition}(Semantic  \cite{Emerson})
A  linear time structure is an element $\mathcal{M}$ in $(2 ^P) ^\mathbb{N}$ . $\forall i  \in \mathbb{N}$,$\forall \mathcal{M} \in (2 ^P) ^\mathbb{N}$:
\begin{itemize}
\item $ (\mathcal{M},i) \vDash p$ with $p \in P$ iff $p \in \mathcal{M}(i)$
\item if $A$ is a propositional combination  of LTL formulas $ (\mathcal{M},i) \vDash   A$  is defined as usual.%%iff  $(\mathcal{M},i) \nvDash A$
\item  $ (\mathcal{M},i) \vDash X(A)$ iff  $ (\mathcal{M},i+1) \vDash A$
\item  $ (\mathcal{M},i) \vDash A U B$ iff  $ \exists j \geq i, (\mathcal{M},j) \vDash B$ and $\forall k ,  i \leq k <j, (\mathcal{M},k) \vDash A  $
\item  $ (\mathcal{\mathcal{M}},i) \vDash A W B$ iff  $\forall j \geq i, (\mathcal{M},j)  \vDash A$ or ( $\exists j \geq i, (\mathcal{M},j) \vDash B$ and $\forall k ,i \leq k <j, (\mathcal{M},k)\vDash A  $)
\end{itemize}

\end{definition}
 Intuitively, the formula $X(A)$ stands for `at the next time $A$ will hold', $AUB$ stands for `$B$ will hold in the future and from current time until $B$ holds, $A$ must hold',
$AWB$ stands for  `if $B$ holds in the future then from current time until $B$ holds, $A$ must hold, and if $B$ will never hold, then $A$ must hold forever(weak until)'. $G(A)$ stands for `at any time $A$ holds' and $F(A)$ stands for `$A$ will hold in the future'. For instance $ \neg i W p $ means that $i$ cannot occur as long as $p$ has not occurred.\\
In the rest of the paper we will assume w.l.g that any LTL formula solely  may contain $\neg$ symbol  applied to propositional variable(s). We call such formula Negative Normal Form (NNF). 

\begin{definition} (LTL SAT problem)
A LTL formula $\phi$ is satisfiable iff there exists a linear model M such that $ (M,0)  \vDash  \phi $.
Conversely, a LTL formula $\phi$ is unsatisfiable iff there is no linear model M such that $ (M,0)  \vDash  \phi $.

%For simplicity we will assume in the rest of the paper that formula is pushed into 
%negative normal form before  solved.
\end{definition}

%\begin{theorem} \cite{GerthPVW95}
%The satisfiability problem of LTL is P-SPACE complete.
%\end{theorem}

\begin{definition} (unsatisfiable core)
An unsatisfiable core of an unsatisfiable formula $ \phi$ is  a formula $\phi'$ such that (1) $\phi'$ is the result of some substitution(s) in $\phi$ of some positive subformula(s)  by TRUE, (2) $\phi'$ has no subformula of the form $A U/W (TRUE)$, $(TRUE) W B$, $A\vee (TRUE)$, $\wedge_i True$ or $X(TRUE)$   and  (3) $\phi'$ still remains unsatisfiable.
\end{definition}
 \begin{table}%
  \begin{center}%
    \caption{LTL-translations of the running example}%
    \label{tab:Cfirst}%
    \subfloat[compliance rules.\label{example}]{\begin{tabular} {|  l|  c|} 
\hline
Rules & LTL\\
\hline
r1.a & $F(o)$\\
\hline
r1.b & $G(\neg c)$\\
\hline
r2.a & $G(o \Rightarrow (F(p)\wedge F(g)))$ \\
\hline
r2.b & $(\neg g) W   p $ \\
\hline
r3.a & $F(i)$\\
\hline
r3.b & $(\neg i) W   p $\\
\hline
r3.c & $G(p \Rightarrow G(\neg i))$    \\
\hline

\end{tabular}}
    \qquad
    \subfloat[unsatisfiable core.\label{core}]{\begin{tabular} {|  l|  c|} %\caption{ Conflicting rules and LTL core}
\hline
Conflicting Rules & LTL core\\
\hline
 & ${\color{green}TRUE} $\\
\hline
 & ${\color{green}TRUE} $\\
\hline
 & ${\color{green}TRUE}$ \\
\hline
 & ${\color{green}TRUE}$ \\
\hline
r3.a & $F(i)$\\
\hline
r3.b & $(\neg i) W   p $\\
\hline
r3.c & $G(p \Rightarrow G(\neg i))$    \\
\hline
\end{tabular}
}
\end{center}
\end{table}
 Table 1.b shows a small unsatisfiable core of our toy example formula.
It is critical  to find a small (or ideally a minimal\footnote{An unsatisfiable core $\phi$ is minimal iff $\phi$ is its only one unsatisfiable core} (MU)) unsatisfiable core in order to detect the cause of a conflict.\\
\begin{theorem}(LTL minimal unsatisfiable core decision problem)\\
Deciding if a LTL formula is a minimal unsatisfiable core is  in P-SPACE %complete

({ \bf sketch of the proof}): \textit{ For each positive subformula of $f$, substitute by $TRUE$ and check unsatisfiability. $f$ is a MU iff any substitution leads to a satisfiable formula. There is a linear number of subformulas, and each checking is in $P-SPACE$. %The hardness comes from a reduction of QBF formula $f_Q$ into a LTL formula $f$. It is shown that  $ f$ is a MU iff $f_Q$ is a MU, which is a P-SPACE hard problem \cite{BuningZ06}.
}
\end{theorem}\\
We furthermore conjecture that the above problem is P-SPACE complete.\\
\cite{Schuppan10} discusses the notion of granularity of core. A coarse unsatisfiable core of a formula $f : f_1 \wedge f_2...\wedge f_n $ only substitutes 
$TRUE$ at the $f_j$ and not in a deeper subformula. Structure preserving translations of the LTL formula $f$ into definitional conjunctive normal form  provide an \emph{equi-satisfiable} formula $f' : f'_1 \wedge f'_2...\wedge f'_m $. The minimal coarse unsatisfiable cores of $f'$ correspond to the minimal unsatisfiable cores of $f$ (see \cite{Schuppan10} for details).  For instance if $f : G(a \wedge \neg b ) \wedge F(b)$, an equi-satisfiable formula may be  $f' : G(x_{a \wedge \neg b}) \wedge G(x_{a \wedge \neg b}\Rightarrow a) \wedge G(x_{a \wedge \neg b}\Rightarrow \neg b) \wedge F(b) $. A coarse MU of $f'$ is $G(x_{a \wedge \neg b}) \wedge TRUE \wedge G(x_{a \wedge \neg b}\Rightarrow \neg b) \wedge F(b) $.  It provides a $f$-MU  : $ G(TRUE \wedge \neg b ) \wedge F(b)$. W.l.g \cite{Schuppan10}, the solver will focus on finding small coarse unsatisfiable core.

\begin{definition}(Closure)
Let $f$ a LTL formula. We note the set of closure variables of $f$ -$ Cl(f)$- as the smallest set  $Set$ such that : 
\begin{itemize}
\item $f \in  Set$
\item If $\psi = \psi_1 \wedge.. \wedge \psi_s \in Set$ and $\psi_j$ is not a conjunction,  $\forall j$  $\psi_j \in Set$ 
\item If $\psi = \psi_1 \vee ..\vee \psi_r  \in Set$ and $\psi_j$ is not a disjunction, $\forall j$  $\psi_j \in Set$
\item If $\psi = F/G(\psi') \in Set$,   $ \psi' \in Set$  and $XF/G(\psi') \in Set$  
%\item If $\psi = G(\psi') \in Set$,   $ \psi' \in Set$  and $XG(\psi') \in Set$  
\item  If  $\psi = \psi' U/W \psi '' \in Set$ , $ \psi''$ and $\psi' \wedge X(\psi)$ are in $Set$
%\item  If  $\psi = \psi' W \psi ''$ , $ \psi'' \in Set$ and $\psi' \wedge X(\psi)$ are in $Set$\\
\item If $\psi = X(\psi') \in Set$ then $\psi' \in Set$
\end{itemize}
Furthermore the number of closure variables of $ Cl(f)$ is linear in the size of $f$ \cite{FischerL79}.
\end{definition}

A traditional mathematical tool to analyze satisfiability is tableau. It is a particular automata of states, whose any state is a subset of $Cl(f)$.
Intuitively, a state is built from a prestate. A prestate is either the starting state containing only the starting formula $f$ either a state containing only closure formulas derived from a precedent state. The derivation of a formula $Xh$ at a state is $h$ at the next prestate. The prestates are intermediary results to build the tableau and do not occur in the tableau except the first one. On Figure 1, the rounded rectangle is a prestate, the others are states. A state is computed by unwinding the formulas and making a choice for the disjunctive ones. For instance, the occurrence of $G( \neg i)$ implies the occurrence of $\neg i$ and $XG(\neg i)$. In Figure 1, at the goal state of transition 1, the   $p$ is chosen  from the disjunction $p \vee (\neg i \wedge X(\neg i W p))$ unwound\footnote{disjunctive unwinding are not shown in the tableau since this is an intermediary result} from $\neg i W p$.

\begin{definition} (state, prestate, $f$-tableau)
The $f$-tableau is a special finite state automata $(St, s_0 ,  R)$ with $St$ the set of states, $s_0$ the initial state and $R \subset ST^2$ the set of transitions. The $f$-tableau is the 'minimal'  automata $A$ such that:
\begin{itemize}
\item Any state of $A$ is a subset of $Cl(f)$.
\item $s_0$ is a prestate  with $s_0 =\cup_i\{ f_i \}$ where $f= \wedge_i f_i$.
\item Let a set $S$  derived from a prestate $PS$ st. $PS \subset S \subset Cl(f) $ and $ \exists \rho$ a total choice function  from $S \cap ( disjunction \cup Future \cup Until \cup WUntil )$ to $S$.  Furthermore, if $S$ is the smallest set $Set$ such that

\begin{itemize}
\item $PS \subseteq  Set$
\item If $\psi = \psi_1 \wedge.. \wedge \psi_s$ and $\psi_j$ is not a conjunction,  $\forall j$  $\psi_j \in Set$ 
\item If $\psi = \psi_1 \vee ..\vee \psi_r $ and $\psi_j$ is not a disjunction,   $ \rho(\psi)= \psi_j \in Set$ for some $j$ 
\item If $\psi = F(\psi')$,   $\rho(\psi) \in \{ \psi';XF(\psi') \}  \cap Set$ 
\item If $\psi = G(\psi')$,   $ \psi' \in Set$  and $XG(\psi') \in Set$  
\item  If  $\psi = \psi' U/W \psi ''$ , $\rho(\psi) \in  \{\psi'' ; \psi' \wedge X(\psi) \}  \cap Set$,

\end{itemize}
 then $S$ is a state of $A$.

\item  Let a set $PS$ containing only all formulas derived from a precedent state $S$ such that  $PS = \{\phi , st. X\phi \in S\}$. Then, $PS$ is  a prestate of $A$.
% and such that if $h= (\rho(g),disj_{\rho(g}) \in UCS(PS)$ then $\rho(h)$ is defined. 
 \item Transitions $R$ of a $f$-tableau stand for the collapsing of $S_1\rightarrow PS \rightarrow S_2$ derivation sequences, i.e., collapsed transitions of the form $S_1 \rightarrow S_2$. 
\end{itemize}
\end{definition}
\begin{figure}

\end{figure}

\begin{figure}[hbt]

\begin{tikzpicture}[shorten >=2pt,node distance=3cm,auto,  every node/.style={draw}]

\node[transform shape ,rounded corners,scale=0.7](q_0) {\begin{tabular}{l}$F(i)$, $ \neg i W p$\\$G(\neg c), \neg c ,X(G(\neg c))$ \\$G(p \Rightarrow G( \neg i))$ \\ \end{tabular}};
\node(q_1) [below of= q_0, scale=0.7,yshift=0.5cm, xshift=0.5cm ] {\begin{tabular}{l}$F(i)$,  $X(F(i))$\\$G(\neg c), \neg c ,X(G(\neg c))$ , $ \neg i W p$ , $p$ \\$G(p \Rightarrow G(\neg i))$, $\neg p \vee G(\neg i)$\\ $G(\neg i)$, $\neg  i $, $XG(\neg i)$  \\$XG(p \Rightarrow G(\neg i))$ \\ \end{tabular}};

%\node(q_2) [ right of= q_0, scale=0.7] {\begin{tabular}{l}$F(i)$, $i\vee X(F(i))$, $X(F(i))$ \\ $ \neg i W p$, $p \vee (\neg i \wedge X(\neg i W p))$,\\$\neg i \wedge X(\neg i W p)$,$\neg i$, $X(\neg i W p)$  \\$G(p \Rightarrow G(i))$, $\neg p \vee G(\neg i)$,\\ $G(\neg i)$, $\neg  i $, $XG(\neg i)$,$XG(p \Rightarrow G(\neg i))$   \\ \end{tabular}};

\node(q_3) [below left of= q_1, scale=0.7, xshift=0.3cm, yshift=-1cm] {\begin{tabular}{l}$F(i)$,  $X(F(i))$ \\$G(\neg c), \neg c ,X(G(\neg c))$ \\ $G(\neg i)$,$ \neg i$ , $XG(\neg i)$,\\ $G(p \Rightarrow G(i))$, $\neg p \vee G(\neg i)$,\\$XG(p \Rightarrow G(\neg i))$ \\  \end{tabular}};

\node(q_4) [below right of= q_1, scale=0.7, xshift=-0.3cm,yshift=-1cm] {\begin{tabular}{l} $F(i)$, $X(F(i))$ \\$G(\neg c), \neg c ,X(G(\neg c))$\\$G(\neg i)$,$ \neg i$ , $XG(\neg i)$, \\$G(p \Rightarrow G(i))$, $\neg p \vee G(\neg i)$,\\ $\neg p$  , $XG(p \Rightarrow G(\neg i))$ \\ \end{tabular}};

\node(q_5)[left of= q_0,scale=0.7,xshift=-1.5cm]  {\begin{tabular}{l}$F(i)$,  $X(F(i))$ \\$G(\neg c), \neg c ,X(G(\neg c))$, $ \neg i W p$,  \\$\neg i \wedge X(\neg i W p))$, $X(\neg i W p)$\\$G(p \Rightarrow G(\neg i))$, $\neg p \vee G(\neg i)$\\ $G(\neg i)$, $\neg  i $, $XG(\neg i)$  ,$XG(p \Rightarrow G(\neg i))$ \\ \end{tabular}};

\node(q_6)[right of= q_0,scale=0.7,xshift=1cm]  {\begin{tabular}{l}$F(i)$,  $X(F(i))$ \\$G(\neg c), \neg c ,X(G(\neg c))$, $ \neg i W p$, \\$\neg i \wedge X(\neg i W p))$, $X(\neg i W p)$\\$G(p \Rightarrow G(\neg i))$, $\neg p \vee G(\neg i)$, $\neg p $ \\$XG(p \Rightarrow G(\neg i))$ \\ \end{tabular}};

\node(q_7)[below of= q_5,scale=0.7, yshift=1cm, xshift=-0.5cm ]  {\begin{tabular}{l}$F(i)$, $X(F(i))$ \\$G(\neg c), \neg c ,X(G(\neg c))$, $ \neg i W p$, \\$\neg i \wedge X(\neg i W p))$, $X(\neg i W p)$\\$G(p \Rightarrow G(\neg i))$, $\neg p \vee G(\neg i)$,$\neg p$\\ $G(\neg i)$, $\neg  i $, $XG(\neg i)$  ,$XG(p \Rightarrow G(\neg i))$ \\ \end{tabular}};

\path[->] (q_0) edge node[transform shape, midway, left , draw=none] {1 } (q_1)
%(q_0) edge (q_2)
(q_1) edge node[transform shape, midway, left , draw=none] {2 }  (q_3) 
(q_5) edge [bend right] node[transform shape, left, draw=none] {14 }(q_7)
(q_7) edge [bend right] node[transform shape, midway, right, draw=none] {15 } (q_5)
(q_1) edge  node[transform shape, midway, right , draw=none] {7 }(q_4)
(q_4) edge [loop below] node[transform shape, ,above, draw=none] {5 } (q_3)
%(q_4) edge [loop below] node[transform shape,  left , draw=none] {6 } (q_7)
(q_6) edge  node[transform shape,near end, above, draw=none] {10 }(q_1)
(q_7) edge  node[transform shape, midway, above, draw=none] {16 }(q_1)
(q_5) edge node[transform shape, near end, above, draw=none] {12 } (q_1)
(q_4) edge [bend right]  node[transform shape, midway, above , draw=none] {6}(q_3)
(q_6) edge [bend right] node[transform shape, midway, above, draw=none] {11 } (q_5)
%(q_6) edge   (q_7)
(q_3) edge [bend right] node[transform shape, midway,  draw=none] {4 } (q_4)
(q_3) edge [loop below] node[transform shape, above  , draw=none] {3 } (q_4)
(q_6) edge [loop below] node[transform shape, midway, above, draw=none] {9 } (q_5)
(q_5) edge [loop above] node[transform shape, midway, above, draw=none] {13 } (q_6)
 (q_7) edge [loop below] node[transform shape, midway, above, draw=none] {17 } (q_7)
;
\draw [<-] (q_0) -- (0,1);
\draw [->] (q_0) -- node[transform shape, midway, above, draw=none] {18 }(q_5);
\draw [->](q_0) -- node[transform shape, midway, above, draw=none] {8 } (q_6);
\end{tikzpicture}

\caption{ Depth-first-search }
\label{dfs}
\end{figure}
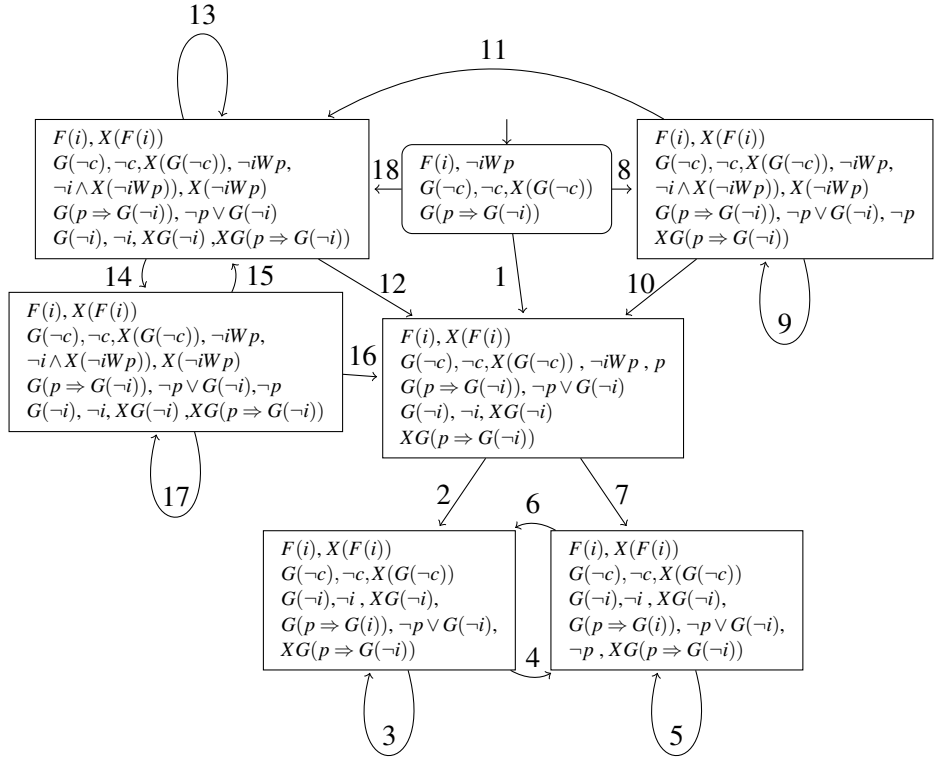

 \begin{theorem}(\cite{KestenMMP93},\cite{GerthPVW95})
 A LTL formula $f$ is satisfied iff there exists a  path of states in the $f$-tableau (finite with no successor at the last state or infinite) starting from the starting prestate and such that any occurrence   of Future  and Until modal operator  in a state of the path fulfills its corresponding promise operand later (in the future) in the path. We call the path : fair path. 
\end{theorem}

In Figure 1, $f$ is a simpler version of our toy example, and there is only unsatisfiable  paths (infinite in this case) since each possible path contains  a Future $F(i)$ but does not realize the promise operand $i$. An argument is that any infinite path will reach in the future a Strongly Connected Component (SCC) where the path will remain in forever. Then $f$ is unsatisfiable.
On-the-fly techniques for satisfiability of temporal logic (eg. \cite{GerthPVW95}, \cite{KestenMMP93}) use nested deep-first-search of fair loop or simple deep first search of fair  SCC.  
 
\begin{theorem}(\cite{Tarjan72},\cite{KestenMMP93})
There exists a depth-first-search algorithm for computing SCCs of a $f$-tableau, and for deciding their fairness.
\end{theorem}

In Figure 1, the exploration steps of simple depth-first-search follow the numbered labels on the transitions. An example of a SCC is the set of states as a support for the set of transitions $\{3 ;4 ; 5 ;6   \}$.

\section{Technical Preliminaries}

We will show how it is possible, by handling conflicts, to enhance above depth-first search method and to drastically shrink the search space. Our solver shown in Section 4 is based on the following intuitions. First, the idea is to record  which occurrences of elements of the closure at a given state entail another one by using unit rule propagation technique from SAT-Solvers.  It enables to extract cause of conflict and non-chronologically backtrack at the last involved choice and eventually to learn information from the conflict, in order to not revisit the same conflict. Furthermore, our solver uses fair prime implicant search to also shrink good but redundant search space. These optimizations enable to only explore the tableau of Figure 2 to decide unsatisfiability of the  running  example $f-tableau$ of figure 1. 
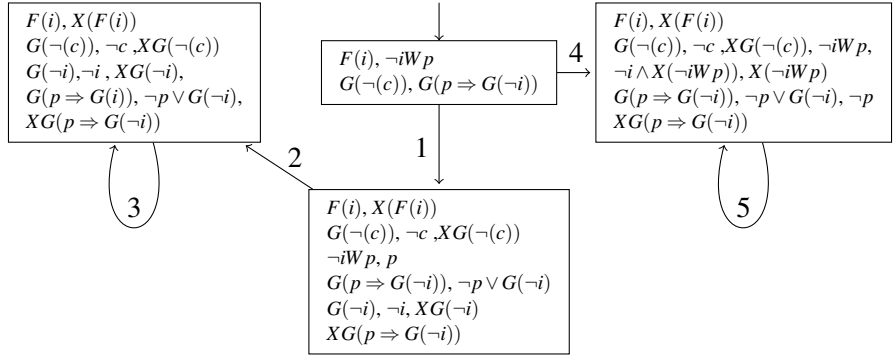
\begin{figure}

\begin{tikzpicture}[shorten >=2pt,node distance=3cm,auto,  every node/.style={draw}]

\node[rectangle,scale=0.7](q_0) {\begin{tabular}{l}$F(i)$, $ \neg i W p$\\$ G(\neg(c))$, $G(p \Rightarrow G( \neg i))$ \\ \end{tabular}};
\node(q_1) [below of= q_0, scale=0.7,yshift=0.5cm ] {\begin{tabular}{l}$F(i)$, $X(F(i))$ \\ $ G(\neg(c))$, $\neg c$ ,$XG(\neg(c)) $ \\ $ \neg i W p$, $p$ \\$G(p \Rightarrow G(\neg i))$, $\neg p \vee G(\neg i)$\\ $G(\neg i)$, $\neg  i $, $XG(\neg i)$  \\$XG(p \Rightarrow G(\neg i))$ \\ \end{tabular}};

%\node(q_2) [ right of= q_0, scale=0.7] {\begin{tabular}{l}$F(i)$, $i\vee X(F(i))$, $X(F(i))$ \\ $ \neg i W p$, $p \vee (\neg i \wedge X(\neg i W p))$,\\$\neg i \wedge X(\neg i W p)$,$\neg i$, $X(\neg i W p)$  \\$G(p \Rightarrow G(i))$, $\neg p \vee G(\neg i)$,\\ $G(\neg i)$, $\neg  i $, $XG(\neg i)$,$XG(p \Rightarrow G(\neg i))$   \\ \end{tabular}};

\node(q_3) [ left of= q_0, scale=0.7, xshift=-1.5cm] {\begin{tabular}{l}$F(i)$,  $X(F(i))$ \\ $ G(\neg(c))$, $\neg c$ ,$XG(\neg(c)) $\\ $G(\neg i)$,$ \neg i$ , $XG(\neg i)$,\\ $G(p \Rightarrow G(i))$, $\neg p \vee G(\neg i)$,\\$XG(p \Rightarrow G(\neg i))$ \\  \end{tabular}};

\node(q_6)[right of= q_0,scale=0.7,xshift=1.5cm]  {\begin{tabular}{l}$F(i)$,  $X(F(i))$ \\ $ G(\neg(c))$, $\neg c$ ,$XG(\neg(c)) $, $ \neg i W p$, \\$\neg i \wedge X(\neg i W p))$, $X(\neg i W p)$\\$G(p \Rightarrow G(\neg i))$, $\neg p \vee G(\neg i)$, $\neg p $ \\$XG(p \Rightarrow G(\neg i))$ \\ \end{tabular}};

\path[->] (q_0) edge node[transform shape, midway, left , draw=none] {1 } (q_1)
%(q_0) edge (q_2)
(q_1) edge node[transform shape, near start, above , draw=none] {2 } (q_3)
%(q_5) edge [bend right] (q_7)
%(q_7) edge [bend right] (q_5)
%(q_1) edge (q_4)
(q_3) edge [loop below] node[transform shape, midway, above , draw=none] {3 } (q_4)
%(q_4) edge [loop below] (q_7)
%(q_6) edge  (q_1)
%(q_7) edge  (q_1)
%(q_5) edge  (q_1)
%(q_4) edge [bend right]  (q_3)
%(q_6) edge [bend right]  (q_5)
%(q_6) edge   (q_7)
%(q_3) edge [bend right] (q_4)
%(q_3) edge [loop below] (q_4)
%(q_6) edge [loop below] (q_5)
(q_6) edge [loop below]  node[transform shape, midway, above , draw=none] {5 }(q_6)
 %(q_7) edge [loop below] (q_7)
;
\draw [<-] (q_0) -- (0,1);
%\draw [->] (q_0) -- (q_5);
\draw [->](q_0) edge node[transform shape, midway, above , draw=none] {4 } (q_6);
\end{tikzpicture}

\caption{ Depth-first-search with conflict handling and prime implicant}
\label{Solver}
\end{figure}
 In the following we explain how unwinding from prestate to state is simulated by  a boolean SAT-problem.

\subsection{From prestate to state: a propositional SAT problem } To tackle the particular choice function handling at Definition 5 (one literal per occurring disjunction) we need a 'three-values' logic which enables partial instantiation. It is also convenient for prime implicant handling. 

\begin{definition}(`three-value' logic, closure variables, literals, clause)
Let $S$ be a set of  LTL formulas.  We call state closure  of $S$ -$ StCl(S)$- any formula met in $Set$ of  the closure algorithm with the initial condition on $Set= S$ instead of $Set=\{f\}$ and without the last rules ($Xg$ derives $g$). For any  element $g$ in the closure  we note $x_g$ a fresh boolean variable, that we call closure variable . This means presence of $g$ in the state. We will use the word `literal' for  $x_g$ or  $\neg x_g$.  Finally, we  call a clause a disjunction of such literals (also represented by a set of literals). 
Let $S' \subset Cl(f)$. We say that $S'$ is conflicting if there exists $h$ and $ \neg h$ in $S'$. Let $V$ be a set of  closure variables, $L$ the literals of $V$.  Then if  $g$ and $h$ are `three-values' logic formulas then $x_{h '} \in V$, $g \wedge h$, and $\neg g$ are `three-values' formulas. Furthermore assuming $S'$ is non-conflicting :
\begin{itemize}
\item $S' \vDash  x_{h'}$ iff $ h' \in S'$ 
\item$S' \vDash  g \wedge h $ iff $ S' \vDash g$ and $ S' \vDash h$ 
\item$ S' \vDash \neg g$ iff $ S' \nvDash g$  
\end{itemize}
We say that a three-values formula  $g$ is valid iff for any non-conflicting set of  $S' $, $S' \vDash  g$. We say that $g $ is fair-valid if for any $S'$ which is a state from any fair path  $S' \vDash  g$.
\end{definition}

\begin{definition}(\label{unwinding} Unwinding clauses from a prestate) 
Let $PS$ a prestate and $Presence(PS)=\{x_{h}|h \in PS\}$. The corresponding Unwound Clause Set $UCS(PS)$ is a set containing the unwound clauses and $AUX(PS)$ the three values conditions. $Set $, $AUX(PS)$ and $UCS(PS)$ are the smallest sets following the rules :

\begin{itemize}
\item $Presence(PS) \subseteq Set \cap UCS(PS)$
\item If $x_\psi = x_{\psi_1 \wedge.. \wedge \psi_s} \in Set$ and any $x_{\psi_j}$ is not a conjunction,\\   $\forall j$ the formulas $x_\psi \Rightarrow x_{\psi_j}\in UCS$ and $\forall j $ $x_{\psi_j} \in Set$ 

\item If $x_\psi = x_{\psi_1 \vee ..\vee \psi_r} \in Set $ and  any $x_{\psi_j}$ is not a disjunction,\\    $x_\psi \Rightarrow (x_{\psi_1} \vee ...\vee  x_{\psi_r})\in UCS$ and $\forall j $ $x_{\psi_j}\in Set$

%\item If $x_\psi = F(x_{\psi'})$,\\   $x_{\psi} \Rightarrow  x_{\psi'} \vee X(x_{\psi}) \in Set$ to the UCS and add  $ x_{\psi'}$ to Set  

%\item If $x_\psi = G(x_{\psi'})$ ,\\ add $x_\psi \Rightarrow  x_{\psi'}$ and $x_\psi \Rightarrow  X(x_\psi) \in Set$ to the UCS and add  $ x_{\psi'}$ to Set  

\item  If  $x_\psi = x_{\psi'} U/W x_{\psi ''}$ ,\\  $x_\psi \Rightarrow  (x_{\psi''} \vee (x_{\psi' \wedge X(\psi)})) \in  UCS$ and $ x_{\psi''}$ and $x_{\psi' \wedge X(\psi)} \in Set$

%\item  If  $x_\psi = x_{\psi'} W x_{\psi ''}$ ,\\ add $x_\psi \Rightarrow  (x_{\psi''} \vee (x_{\psi'} \wedge X(x_\psi))) \in Set$ to UCS and $ x_{\psi''}$ and $x_{\psi'} \wedge X(x_\psi)$ to $Set$\\

\item $x_h,x_{\neg h} \in Set $   then $\neg x_h \vee \neg x_{\neg h} \in AUX$

\end{itemize}
\end{definition}

Furthermore, $AUX(f)$ (resp. $UCS(f)$, $ Presence(f) $)is the union of $AUX(PS)$ for any $PS$ in the $f$-tableau (resp $UCS(PS)$, $Presence(PS)$). 
 The unwound formulas $UCS(PS) \setminus Presence(PS)$ are fair valid formulas (see proof section 5) and  of the form $x_\phi \Rightarrow disj_{x_\phi}$ where $disj_{x_\phi}$ is the classical disjunctive unwinding of closure formulas \cite{GerthPVW95}, \cite{KestenMMP93}.

 The formula $f$ of Figure 3 provides the clause $UCS(f)$ :

\begin{tabular}{l|r}

 $x_{F(i)} \Rightarrow x_i  \vee x_{XF(i)}$ & $x_{ G\neg c )} \Rightarrow x_{XG\neg c }$\\
 $x_{ G\neg c } \Rightarrow x_{\neg c }$ & $x_{(\neg i) W p} \Rightarrow x_{p } \vee x_{\neg i \wedge X((\neg i) W p)}$\\
  $x_{\neg i \wedge X((\neg i) W p)} \Rightarrow x_{\neg i } $ &  $x_{\neg i \wedge X((\neg i) W p)} \Rightarrow x_{X((\neg i) W p)} $\\

 $ x_{G(p \Rightarrow G(\neg i)))} \Rightarrow x_{XG(p \Rightarrow G(\neg i))}$ &  $x_{G(p \Rightarrow G(\neg i)))} \Rightarrow x_{p \Rightarrow G(\neg i)}$\\
$x_{p \Rightarrow G(\neg i)} \Rightarrow x_{\neg p} \vee x_{G(\neg i)} $ &  $x_{G(\neg i)} \Rightarrow x_{XG(\neg i)} $\\
  $x_{G(\neg i)}\Rightarrow x_{\neg i}$ & $\forall v \in$ $CLSTf$ $ \neg x_v \vee \neg x_{\neg v}  $\\
\end{tabular}

\begin{proposition}
An instance $IS$ of the boolean SAT problem $UCS(PS) \cup AUX(PS)$ provides a state $S$ from $PS$ and reciprocally.
\end{proposition}

Since many instances correspond to a state in the tableau, and since several states may be redundant regarding LTL satisfiability problem, we introduce Fair Prime Implicant. 
 
\begin{definition} (Fair Prime Implicant)
Let $IS$ as above, a Fair Prime Implicant $IS.FPI$ of $IS$  is a maximal\footnote{Intuitively the switching simulates the removal of closure element in corresponding state } switching from some  assigned $ x_h$ at $IS$ to $\neg x_h$ such that $h$ is not a promise operand and $IS.FPI \vDash UCS(PS) \cup AUX(PS)$. At a given $IS.FPI$ it corresponds only one state $FPI$ in the $f$-tableau.
\end{definition}

\begin{theorem}(Fair prime implicant version of  Depth-First-Search )
A formula $f$ in LTL is satisfiable iff there exists a fair path solely with FPIs as states.(\emph{ proof is omitted }).\\
%\begin{itemize}\
%\item If $f$ is satisfiable, then there exists a fair path $p$. It is possible to extract from the first state a  FPI, by incrementally trying to delete an (non operand of promise) element and correspondingly changing the function choice $\rho$. The same process can be done at any state of $p$. It provides a new path $p'$ of FPIs. Furthermore, if $p'$ contains a promise at the $i^{th}$  FPI, $p$ also contains at the corresponding $i^{th}$ state. Since $p$ is fair,   there exists a state at time $i'\geq i$ containing the operand of the promise. But by construction the $i'^{th}$ FPI of $p'$ contains any operand of promise and particularly the one introduced above.
%\item The reciprocal is obvious 
%\end{itemize}
\end{theorem}\\
For instance, the FPI technique enables in our depth first search to  ignore the goal state of the transition number 4 at Figure 1.

To solve the boolean SAT-problem current solvers use unit rule propagation \cite{Davis1962}.\\
\begin{definition} (Unit rule propagation)

\begin{itemize}
\item Each instantiated literal must be propagated\footnote{a Weakest version and optimized one  of current solvers requires only propagation along watched literals \cite{MoskewiczMZZM01}} over any non yet satisfied clause containing the opposite one. This opposite literal is then temporally erased from the clause. 

\item If a clause becomes unit literal $l$ because of unit rule propagation(s), then $l$ is assigned

\end{itemize}
\end{definition}

This propagation is critical for conflict analysis.
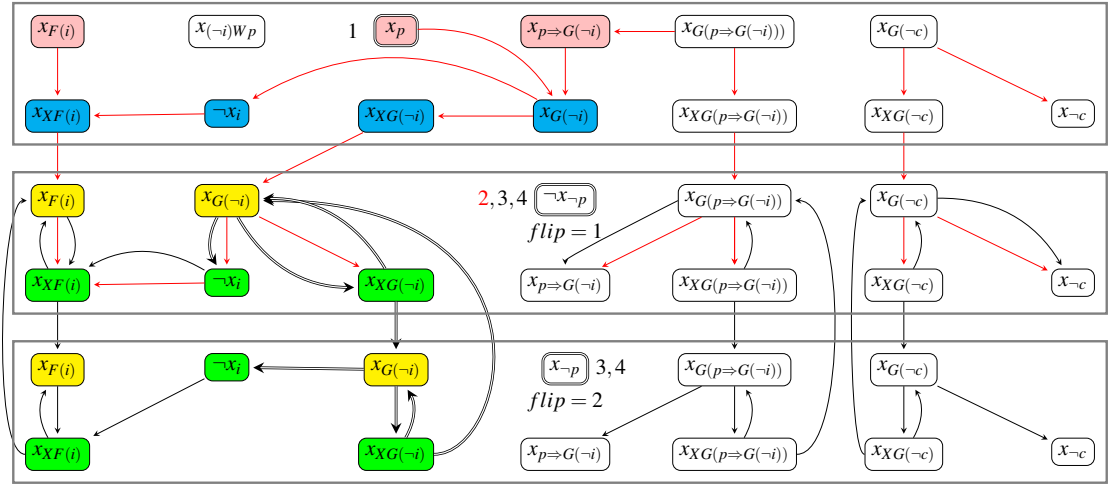
\begin{figure}
\caption{Implication graph and conflict analyses}
\scalebox{0.75}{
\begin{tikzpicture}[>=stealth,->,shorten >=2pt,auto]
\matrix [matrix of math nodes,
column sep={3
cm,between origins},
row sep={1.5cm,between origins},
nodes={ rectangle ,rounded corners, draw,minimum width=7.5 mm},ampersand replacement=\&]
{ |  (A1) [fill=pink]  |   x_{F(i)}    \& |(B1)|  x_{(\neg i) W p}   \&   | (D1)[double, fill = pink,label={ left :1}]|  x_{p}  \& | (C2)[fill=pink]|
  x_{p \Rightarrow G(\neg i)}  \&      | (C1)|   x_{G(p \Rightarrow G(\neg i))) } \& | (U1) |  x_{G(\neg c)}  \&  \\ 
| (E1) [fill=cyan] |  x_{XF(i)}  \& |(F1) [fill=cyan]|  \neg x_{ i}  \& |(G1)[fill=cyan]  |  x_{XG(\neg i)}  \& | (D2) [fill=cyan]|  x_{G(\neg i)}   \& |(Y1) |  x_{XG(p \Rightarrow G(\neg i))}  \& | (U2) |  x_{X G(\neg c)}  \& | (V2) | x_{ \neg c}    \\ 
 |  (A) [fill=yellow] |    x_{F(i)}    \& |(B)[fill=yellow] |  x_{G(\neg i)}  \&   \& 
|(D) [double,label={ left :\textcolor{red}{2},3,4},label={below : flip=1}] | \neg x_{\neg p}  \&      | (C)|   x_{G(p \Rightarrow G(\neg i))}   \& | (U3) |  x_{G(\neg c)}  \& \\ 
| (E) [fill=green] |  x_{XF(i)}  \& |(F) [fill=green]|  \neg x_{i}  \& |(G) [fill=green]|  x_{XG(\neg i)}  \&|(T2)| x_{p \Rightarrow G(\neg i)}  \& |(Y) |  x_{XG(p \Rightarrow G(\neg i)) }  \& | (U4) |  x_{X G(\neg c)}  \& | (V4) |  x_{\neg c}    \\ 
| (I) [fill=yellow] |  x_{F(i)}  \& |(J) [fill=green]|  \neg x_{ i}  \& |(K) [fill=yellow] |  x_{G(\neg i )} \& |(Z1)[double,label={ right :3,4},label={below : flip=2}]|  x_{\neg p}\& |(Z) | x_{G(p \Rightarrow G(\neg i))}    \& | (U5) |  x_{G(\neg c)}  \& \\ 
| (M) [fill=green]|  x_{XF(i)}  \&  \& |(O) [fill=green]|  x_{XG(\neg i)}   \& |(T1)|  x_{p \Rightarrow G(\neg i)}  \& |(W) |  x_{XG(p \Rightarrow G(\neg i))}   \& | (U6) |  x_{X G(\neg c)}  \& | (V6) |  x_{\neg c }   \\ };
\begin{scope}[every node/.style={font=\small\itshape}]
%\draw (0,2.5) -- (A) ;
%\draw  (A) --  node[near end ][left] [yshift=0.2cm]{1}(B) ;
%\draw [dotted](B) --  node[near end ] [left]{n}(E) ;
%\draw[dashed] [double] (B) to[bend right]  (A) ;
\draw [red] (A) --  (E) ;
\draw  (A) edge [bend left] (E) ;
\draw  (Z) --  (T1) ;
\draw [red] (C) --  (T2) ;
\draw  (C) ..controls +(left:1cm) and +(up:0.5cm)..  (T2) ;
\draw  [red](Y1) --  (C) ;
\draw [red] (A1) --  (E1) ;
\draw  [red](E1) --  (A) ;
\draw  [red](F1) --  (E1) ;
\draw  [red](G1) --  (B) ;
\draw  [red](D1) edge [bend left]  (D2) ;
\draw  [red](C2) --  (D2) ;
\draw  [red](C1) --  (Y1) ;
\draw  [red](D2) --  (G1) ;
\draw  [red] (U1) --  (U2) ;
\draw [red]  (U1) --  (V2) ;
\draw  [red] (U3) --  (U4) ;
\draw  [red] (U3) --  (V4) ;
\draw  (U5) --  (U6) ;
\draw  (U5) --  (V6) ;
\draw  [red](U2) --  (U3) ;
\draw  (U4) --  (U5) ;
\draw  (U6) edge [bend right]  (U5) ;
\draw  (U6)  ..controls +(left:1cm) and +(left:1cm)..   (U3) ;
\draw  (U3) edge [bend left]  (V4) ;
\draw  (U4) edge[bend right]  (U3) ;
\draw [red] (D2) edge[bend right] (F1) ;
\draw  [red](C1) edge[above] (C2) ;
%\draw (C) --  node[near end ] {5} (E) ;
\draw [red](F) --   (E) ;
\draw (F) edge [bend right]   (E) ;
\draw (E) --  (I) ;
\draw (O)  edge [double,bend right]  (K) ;
\draw  (I) --  (M) ;
\draw (M) ..controls +(left:1.2cm) and +(left:1cm)..  (A) ;
\draw [double](O) ..controls +(right:2cm) and +(right:6cm)..  (B) ;
%\draw  (P) ..controls +(right:1cm) and +(right:2cm)..  (D) ;
\draw (W) ..controls +(right:2cm) and +(right:2cm)..  (C) ;
%\draw (Y) ..controls -(left:1cm) and -(left:0.5cm)..  (C) ;
\draw [double]  (K) edge  [double](J) ;
\draw [double](B) edge [double,bend right]  (G) ;
\draw [red](B) --  (G) ;
%\draw (L) --  (P) ;
%\draw [double](Z) --  (P) ;
\draw [red](B) --  (F) ;
\draw (B) edge [double, bend right]  (F) ;
%\draw (D) --  (H);
%\draw (C) -- (H) ;
\draw [red](C) -- (Y) ;
\draw [red](C) -- (Y) ;
\draw  (Y) edge [bend right] (C) ;
\draw  (W) edge [bend right] (Z) ;
\draw  (E) edge [bend left] (A) ;
\draw  (M) edge [bend left] (I) ;
\draw    (G) edge [double,bend right] (B) ;
\draw (Y) -- (Z) ;
\draw   (J) -- (M) ;
\draw (G) edge [double] (K) ;
\draw  (K) edge [double] (O) ;
%\draw (Z) -- (P) ;
%\draw (H) -- (L) ;
\draw (Z) -- (W) ;
\end{scope}
\node[draw=gray,inner sep=6pt, very thick,rectangle,fit=(A1) (B1) (C1)  (D1) (C2) (E1) (F1) (G1) (D2) (U1) (U2) (Y1) (V2)  ] (a){};
\node[draw=gray,inner sep=6pt, very thick,rectangle,fit=(A) (B) (C)  (E) (F) (G) (Y) (U4) (U3) (V4)  ] (a){};
\node[draw=gray,inner sep=6pt, very thick,rectangle,fit=(I) (J) (K) (Z)  (M) (O) (W)  (U5) (U6) (V6)] (b) {};
%\draw[very thick] (a) -- (b)  {};
\end{tikzpicture}}
\end{figure}
In the following we show how to handle  unit rule propagations to support conflict analyses.

\subsection{Implication Graph to support  Conflict analyses } 
The Implication Graph is an extension of propositional SAT-solvers' one to LTL-tableau. The intuition is to record the occurrences of elements of the closure at a given state that entail another one. An Implication Graph is  a bicolor graph $(Nodes,T_{red},T_{black})$ where $T_{red}$ and  $T_{black}$ are subsets of $Nodes^2$. Figure 3 shows a part of the Implication Graph adapted from the $f$-tableau of Figure 1. Intuitively, the Implication Graph is a concatenation of several $IS$s implication graphs denoted $IS.IG$. The red part $T_{red}$ is used for conflict analysis of the depth-first-search stack $S$ and it is a DAG, and the black part $T_{black}$ records some past red edges and corresponds to the conflict analysis of the SCC-search using stack $S'$ and allows loop for inductive reasoning . \\
{ \bf Nodes' feature}  Intuitively, a Node $N$ stands for an assigned literal at a given state. On Figure 3, the rounded corners rectangles are Nodes. Each node is inside a big rectangle standing for state. More precisely, a Node  corresponds to the ongoing prestate, to an ongoing $IS$ while it is found and to a chosen extracted $IS.FPI$ in this case. On Figure 3,  the three  states\footnote{for understanding but w.l.g, the below state is not a FPI}  are the one which support the transitions $ \{ 2 ; 3 ; 4\}  $ on Figure 1. Furthermore a Node can be either choosen or required. On Figure 3, a chosen node  is doubly surrounded. The level of a chosen Node $N$ is its chronological order of choice in the whole $f$-tableau. On figure 3 numbers are levels of chosen nodes. The level of any node $N$ is the maximum\footnote{0 if no ancestor nodes are choosen} level of the choosen nodes which involve $N$ i.e which are ancestors of $N$ in $T_{red}$. The level of a set of nodes is the maximum level its nodes. A required node is either without antecedent but with level $0$ either gets an antecedent in $T_{red}$. \\ 
{ \bf Transitions' feature}   If a Node $N_l$ which corresponding literal $ l $  comes from a clause $C= \vee_j l_j \vee l$ which has become unit, then the red edges
 $(N_{\sim l_j},N_l)$ are in $T_{red}$ just after this unit propagation. Let's focus on the above state. For instance, 
$(x_{p},x_{ G( \neg i)})$ and $(x_{p \Rightarrow G( \neg i)},x_{ G( \neg i)})$  are red edges because of the unit rule from the clause $x_{p \Rightarrow G( \neg i)}\Rightarrow (x_{\neg p} \vee x_{ G( \neg i)}$. Furthermore, the derivation from a state to a next state  is also recorded using red  edges  such that the occurrence of $ x_{Xg} \in IS.FPI $  entails the occurrence of $x_g$ at the    next prestate. For instance on Figure 3, the above FPI derives to the middle one, thus there exists a red edge in the graph from $x_{XF(i)}$  at above state  to $x_{F(i)}$ at the next one.

Furthermore, while a FPI is revisited, then the current $IS$ implication graph $IS.IG$ has to be connected to the first one $IG_{old}$ which visited the same FPI. The algorithm creates   black transitions from any  nodes $N(\neg x_{op_{pro}})$ (resp. $x_{Xh} \in IS.FPI$) at $IS.IG$ to the same literal  one of $IG_{old}$. This connection is called `bind' function.  
 For instance, for corresponding derivation  on Figure 1 for $IS.IG$ at the goal state of transitions $\{ 3; 5 ; 6\}$ the $IS.IG$ and $IG_{old}$ are the same. For simplicity and w.l.g\footnote{The particular computation of fixpoint remains the same while superimposing in this simple case} they have been superimposed at Figure 3. In this case, the  transitions of `bind'have been omitted w.l.g, and  solely the bottom-up edges from source state to the goal state of transitions $\{ 3; 5 ; 6\}$ are shown (e.g.,$x_{XF(i)} $  at the below state to  $x_{F(i)}$ at the same state for transition 3).
Finally, given a $T_{red}$ and a choosen Node $N$ still red, $flip(Nodes,T_{red},T_{black},N) = (Nodes    \cup  \{ \sim N (red)\}, T_{red}\setminus\{(N_1,N_2) \in T_{red} | level(N_2) \geq level (N) \},T_{black}) $   is the flipped  Implication Graph  regarding $N$ with $\sim N (red)$ a fresh node.\\

\section{Solver}
Our depth-first search temporal conflict driven solver is a combination of depth first search of fair SCC in tableau \cite{KestenMMP93} and of boolean SAT-solver. Thus, our solver uses unit rule propagation method,   boolean conflict handling \cite{MoskewiczMZZM01}. It  also uses  a new temporal conflict driven method inspired by resolution for temporal logic\cite{Fisher91}. \\
{ \bf Basic Solver}\footnote{The main components of the algorithm are shown in a recursive form for convenience}
 Algorithm 1 shows the main method of the algorithm called Solver. At each new prestate, the solver  populates by  clauses by unwinding the prestate according to  Definition \ref{unwinding}. Otherwise, unit  rules and boolean conflict detection\footnote{ a boolean conflict detection occurs while a  clause is falsified by current partial assignment} are launched. A Backtrack (Algorithm 2) is triggered in case of a conflict, otherwise if it is possible, a choice of literal following a  heuristic is done. Once an $IS$ is found and a FPI extracted, then  a SCC-search-forward (Algorithm 3) function is called. Otherwise the Solver is recursively called.
      \begin{algorithm} \label{solver}

\caption{Solver }

\begin{multicols}{2}
\SetNoline
 
\uIf{not unwound}  {  Unwind\;} 
 Unit-rule ; bool-conflict-detection \;
		\uIf {conflict} { Backtrack;}
			 \uIf{$IS$ found}  {SCC-search-forward;} 
		\uElse{ make a choice of literal\;  Solver \;}

\end{multicols}
\end{algorithm}

{\bf Propositional Conflict Handling while backtracking} 
 A Propositional Conflict Handling is triggered when a clause is falsified (or equivalently when a literal and its opposite occurs).
Similarly to SAT-solvers'one, the Propositional Conflict Handling starts from a set of conflicting nodes $Nodes_C$ and corresponding literals  $C$ which falsifies the clause $\neg C$ and analyzes which nodes have involved those conflicting literals using $(Nodes(red),T_{red})$. Let $\mathcal{A}(C)$ be the subDAG of $(Nodes(red),T_{red})$ which stands for ancestors of $Nodes_C$. Let $\mathcal{A}(C)(conflict-level)$ be the subDAG of $\mathcal{A}(C)$ with nodes of `conflicting' level of $Nodes_C$ ie. $conflict-level$  and $N(conflict-level)$ the  choosen node of level $conflict-level$ . Let $Limit(C) = \{N(conflict-level)\} \cup (Parent_{T_{red}}[\mathcal{A}(C)(conflict-level) \setminus  \{N(conflict-level)\}] \cap Level(conflict-level-1,\mathcal{A}(C))) $ where $Level(m,\mathcal{A}(C))$ means the subDAG of $\mathcal{A}(C)$ with node level at most $m$. We call limit conflict clause $\neg Limit(C)$. The last conflicting choosen node $N(conflict-level)$ is then switched if the corresponding flipped partial assignment has not been visited yet (node.flip=1). In this first case, similarly to boolean SAT-solvers,   the function `Conflict-require' adds red edges to  $(Nodes(red),T_{red})$  : the red transitions with a source node in $Limit(C) \setminus \{N(conflict-level) \}$ to the goal node $ \sim N(conflict-choice) $ . However, differently from boolean SAT-solver, since the algorithm  records informations in black part $(Nodes(black),T_{black})$  in the second case (flip=2), the same transitions but in color black are added.   Furthermore, $ \sim N(conflict-level) $  is now required and not choosen. Those red or black edges are to ensure we can compute the reason of the requirement of  $\sim N(conflict-level)$. Finally, if the conflict level is 0 then  the algorithm terminates by unsatisfiable. 
\begin{algorithm}\label{Backtrack}
\SetNoline
\begin{multicols}{2}

Compute Conflict-level\;
\uIf{Conflict-level=0}{print (`unsat') , break\;}
State-Conflict-Clause-learning\;
Tableau.IG.erase(Conflict-level)\;
stack-s.erase(C-level)\;
stack-s'.erase(C-level)\;
Conflict-require\;
SCC-search-backward; 
%//}
\end{multicols}
\caption{Backtrack}
\end{algorithm}

On Figure 3, the backtrack is done from the conflicting (see. TC-Analysis) nodes $x_{G(\neg i )}$ and  $x_{F(i)}$ at the middle state. Following the red part, the last involved and chosen node is $x_p$ at above state.  While backtracking  bad states and corresponding nodes are erased ( above  state at Figure 3). On the contrary to propositional SAT-solver, the algorithm has to record the cause of these states to be bad (to avoid revisiting them) using a conflict clause per  state\footnote{We ask that the conflicting clause forbids corresponding red FPI of state}.These learned clauses must not be forgotten.  On figure 3, the yellow literals are conflicting literals at middle state but the clause $\neg x_{G(\neg i)} \vee \neg x_{F(i)}$ has already been learned. At above state, the pink literals provide the learned  clause   $\neg x_{p} \vee \neg x_{p \rightarrow G(\neg i ) } \vee \neg x_{f(i)}$.  
Finally, a SCC-search backward is launched.
  Algorithm 2 summarizes the above ideas.  We refer to \cite{ZhangMMM01} for more details about backtracking in boolean SAT-solvers. 

\begin{algorithm}
\SetNoline
\begin{multicols}{2}
\uIf{$FPI$ is new}
      {Nb(FPI):=i:=i+1\;
Lp(FPI):=Lv(FPI:=Nb(FPI)\;
stack-s.push(FPI)\;stack-s'.push(FPI)\; parent=FPI; prestate=FPI.next()\; Solver\; }
  \uElse{ \uCase {$state \in stack-S$ } 
   {Lp(Parent):= min(Lp(Parent),Nb(FPI));}
  
\uCase{ $ FPI \notin stack-S$ $\wedge$ $Nb (parent) > Nb (FPI)$}{Lv(Parent) :=min(Lv(parent),Nb($FPI$))\;
 parent.unr-prom= parent.unr-prom $\cap$ $FPI_{old}$.unr-prom\;
$bind(IS.IG,IG_{old})$\; 
SCC-search-backward \; }}

\end{multicols}
\caption{SCC-search-forward}\label{scc-forward}
\end{algorithm}

{\bf SCC-Search-Forward} The SCC-search-forward shown Algorithm \ref{scc-forward} is similar to the `forward' part of the computation of strongly connected components and uses depth first search numbers (Lp,Nb,Lv). If the FPI  is new, then  new numbers are computed and if it is possible, the next prestate (and corresponding prestate Nodes and transitions from derivations) are created from the (red) nodes from literals $x_{Xh} \in IS.FPI$, otherwise the problem is satisfiable.  Moreover, if the already visited $FPI_{old}$ is still in $Stack-S'$ or in $Stack-S$,  a computation on Tarjan's numbers\footnote{Please see for more details about Tarjan's numbers \cite{Tarjan72}} is also launched.  The unrealizable promises are also computed. Furthermore, in any revisiting case, a rollback is launched while calling SCC-search-backward (see Algorithm 4).

{\bf SCC-Search-Backward} 
First the algorithm adds black copies of red edges in $IS.IG$. Then, starting with the current choosen node $N$ of current level, the  Algorithm 4 simply finds the last non-flipped chosen node. If it is in $IS$ then, it calls $flip(IG , N)$ and Solver. Otherwise change color red to black at the `next' edges from $parent.IG$ to $IG$. Then a $SCC$ test over Tarjan numbers is launched from the parent state, and if a $SCC$ is found a SCC-handling is called, otherwise, update of unrealizable promise is done. If a promise is unrealizable then SCC-handling  calls a Temporal Conflict Analysis (TC-Analysis), otherwise the problem is satisfiable.

\begin{algorithm}\label{scc-back}
\SetNoline
\dontprintsemicolon
\begin{multicols}{2}

N=node(level)\;
IS.IG.edges.black-copies\;
 \uIf{N.flip=2 $\wedge N \in IS$  } {level=level-1; SCC-Search-Backward}\;
	\uIf {$N.flip=1 \wedge N \in IS$}{ flip(IG,N)  \; Solver;}
\uIf{$N \notin IS$} { red-to-black-parent.IG-IS.IG-derivation \; FPI=parent; pop stack-s\; parent= head stack-s\;
			\uIf{ Lp($FPI$)=Nb($FPI$)=Lv(FPI)}{$SCC-handling^*$}
			 \uElse{ Lp(parent)=min(Lp(parent),Lp(FPI))\; Lv(parent)=min(Lv(parent),Lv(FPI))\;
parent.unr-prom= parent.unr-prom $\cap$ $FPI$.unr-prom\;  SCC-search-backward}}
\end{multicols}
\BlankLine
 $SCC-handling^*$ ::\;
\uIf{$unrealizable promise= \emptyset$}{print `satisfiable'; break; } \uElse {TC-Analysis;  }

\caption{SCC-search-backward}
\end{algorithm}  

 { \bf TC-Analysis of unfair SCC}  In the SCC, the algorithm 5 chooses an unfair promise and computes a backward fixpoint from  some nodes $N(\neg x_{op(Promise)})$ for any SCC states along the recorded black implication graph. Precisely, except the root state of the SCC, any state of the SCC gets a corresponding black `IG' from $stack-S$ which is the $IS.IG.edges.black-copies$ one while SCC-backward-search. For the root state SCC, only the nodes $N(x_{Xh})$ and $N(\neg x_{op(Promise)})$ get some black transitions. 

\begin{algorithm}\label{tconflict}

\SetNoline
\begin{multicols}{2}
 INI: Vector= $\neg ops(Promise)$ $\cap SCC$\;
\While{$\exists e \in Vector  \wedge e$ not marked} { mark e ;
			 $v=e.black-parents $    \;  \lFor {$ l \in  v \wedge l$ not marked } { Vector.push(l)}  				
}
					
$\forall state \in SCC$ pick up a State.IG\; do
 learn( $Vector \cap state.IG.prestate, promise$)\;

 erase SCC\;

 Backtrack\;
\end{multicols}
\caption{ Temporal Conflict Analysis}

\end{algorithm}  
The fixpoint computation starts from those nodes at $IS.IG.edges.black-copies$ or particular nodes at the root. Once the inflationary backward fixpoint using $T_{black}$ is terminated, then at each state in  $SCC $, the algorithm picks up a corresponding\footnote{Since the root has been revisited, it gets at least one black $IG$} $IG$. For any state,  the  `prestate(s)' Nodes $Nodes_{prestate} $  in the $IG$ which are  also in the fixpoint are declared conflicting with the unfair promise and the algorithm learns and must not forget the conflict clause. Then, the method erases all the states of this SCC. It finally triggers  a classical Backtracking at the nodes of the $Root$ from the conflicting prestate(s)Nodes  of  the root. 
At Figure 3, the unfair promise is $F(i)$, and the fixpoint computation is shown by double arrow. In this SCC, the yellow and green Nodes are involved in the temporal conflict, and the yellow are the causes of this conflict, ie., $x_{F(i)}$ and $x_{G(\neg i)}$ are conflicting. Thus, $\neg x_{G(\neg i)} \vee \neg x_{F(i)}$ is learned forever.

\section{Correctness, Completeness, and Extraction of a small unsatisfiable core}
\begin{lemma}\label{aux-valid}
Any clause from $AUX(f)$ or $UCS(f) \setminus Presence(f) $ are fair valid.\\
{ \bf  proof}: Any fair state is non conflicting then AUX is fair valid. By construction, any fair state satisfies any clause from $UCS(PS) \setminus Presence(PS) $.
\end{lemma}

\begin{lemma}\label{next-valid}
Let $f$ be a LTL formula. Assume the Algorithm has computed a conflict analysis from the  conflicting literals $C$. Let $ICl(f,C) = AUX(f) \cup UCS(f) \cup Learn(f,C) $ with $Learn(f,C)$ containing any  learned clause   occurring in the algorithm strictly before $C$ and  any limit conflict clause\footnote{ the limit conflict clause is $\neg Limit(C)$; we consider it even if the limit conflict clause is not learned by the solver} occurring at any conflict handling strictly before $C$.  Assume that  $Learn(f,C)$ are fair valid clauses. Let $Cf$ be the conjunction of conflicting literals used to  learn a resulting  clause of the conflict analysis $\neg Cf$. Then $\neg Cf $ is fair valid.\\
{ \bf sketch of the proof}:
Thanks to lemma \ref{aux-valid}, $ICl(f,C) \setminus Presence(f) $ are fair valid clauses.
 Let any state $S$ from any fair path $p$ of any tableau of a temporal logic formula.  Assume now that $S \vDash Cf$. We have two cases:
\begin{enumerate}
\item Either the conflict $C$ is boolean. Let $\mathcal{A}'(C)= Level(\mathcal{A}(C),conflict-level)$ and $Limit(C)$ as above. Then  each  node in $\mathcal{A}'(C) \setminus  \{n(conflict-level)\}$  is  required and it  originates either from state to prestate derivation, either from a clause $Cl \in ICl(f,C) \setminus Presence(f) $ which has become unit at a given state. Since $ICl(f,C) \setminus Presence(f) $ are assumed fair valid then  $S' \vDash Cl$ for  such a clause $Cl$ and for any  state $S'$  in $p_{S}$\footnote{suffixes of p from $S$}. Then the proof from $\mathcal{A}'(C)$ by unit rule of the conflict $C$ of our algorithm implies that there exists a state $S_{conflict}'$ in $p_{S}$  such that $S_{conflict}'$ contains $\square$. This implies a contradiction since $p$ is assumed fair and then no state of $p$ should be conflicting.
\item Either the conflict $C$ is temporal. Assume $S'$ any  state  of the unfair SCC. For any state of the SCC, let $Pre(S')$ be the  set of `black' prestates  from  a chosen $IS.IG.$ from $S'$. Let $k \in \mathbb{N} $.  Imagine virtually the  exploration of any non conflicting prefix path $p'$ of length k in the induced tableau $T(Pre)$ by also considering $ICl(f,C) \setminus Presence(f) $. It consists of the building of a boolean SAT-problem based on the following observations:

\begin{itemize}
\item Since any bad old SCC is not reachable by not forgetting any conflict clause of bad state/bad SCC, then there exists a  k-depth-first navigation over the Prime Implicants from $T(Pre)$ but remaining in the unfair SCC and following the Prime implicant depth-first-search of the $f$-tableau. 
\item if $(s_{i_0}=Pre,s_{i_1},....,s_{i_k})$ is a Prime Implicant path in $T(Pre)$ from the precedent k-depth-first navigation, then from the algorithm, at any transition $(s_{i_{j}},s_{i_{j+1}})$, it corresponds (several) state(s) Implication graphs $IG_{i_j}$ for  $s_{i_{j}}$, and $IG_{i_{j+1}}$  for   $s_{i_{j+1}}$ corresponding at any (re)visit of the states.
\item There exists a k-depth-first  navigation of  full $T(Pre)$ following the Prime implicant depth-first-search of the $f$-tableau, such that if $(s'_{i_0}=Pre,s'_{i_1},....,s'_{i_k})$ is a path of states in $T(Pre)$, then a corresponding Prime implicant path  $(s_{i_0}=Pre,s_{i_1},....,s_{i_k})$ is one from  k-depth-first navigation of the Prime implicant $f$-tableau.
\item  Let $Cl_0(Pre)$, $UC_k(f)$ $\setminus Presence_k(f)$ , $AUX_k(f),Learn_k(f,C)$ be the timestamped variables and corresponding clauses. Let  $Next_k = \{ x_j(X(f)) \Rightarrow x_{j+1}(f)$$|$$0 \leq j < k$  $\}$ be the clauses encoding the state to next prestate derivations. Then there exists a DLL-exploration $E $ of the propositional problem $Cl_0(Pre) \cup UC_k(f)$$ \setminus Presence_k(f) \cup AUX_k(f) \cup Learn_k(f,C) \cup Next_k$ following the k-depth-first navigation of the full $T(Pre)$ but disregarding conflicts which do not occur in the DFS of the SCC in the $f$-tableau.
\item Let $E'$ be the modified exploration of $E$ but by pruning any part of the exploration which contradict any timestamped limit conflict clause.
\item Let $E_{Promise}$ be the modified exploration of $E'$ for the boolean SAT problem $Cl_0(Pre)$, $UC_k(f) $ $\setminus Presence_k(f)$ , $AUX_k(f),Learn_k(f,C),Next_k,x_k(Promise)$ without learning. Furthermore it non chronologically backtracks. It also  considers only conflicts of the form $\{ x_k(op(Promise))$ $  ; \neg x_k(op(Promise))\}$. Then clearly $E_{Promise}$ does not find any solution because the promise is not fulfilled and particularly at step k, ie. the boolean problem is unsatisfiable.
\end{itemize} 

 It is now feasible to show that :
\begin{enumerate}
\item The last conflict $C_{last}$ of $E_{Promise}$ is at level 0. This means that ancestor  literals in $\mathcal{A}_k (C_{last})$ with no parent gets a level 0, ie.  they correspond to clauses $Core_k$ of length one in $Cl_0(Pre)$, $UC_k(f) \setminus Presence_k(f)$ , $AUX_k(f),Learn_k(f,C),Next_k,x_k(op(Promise))$ since there is no learning in $E_{Promise}$.  Furthermore $x_k((op(Promise)) \in Core_k$. Finally, $Core_k$, $UC_k(f) \setminus Presence_k(f)$ , $AUX_k(f),Learn_k(f,C),Next_k,x_k(op(Promise))  $ is an unsatisfiable core.
\item   Let $Cf'_k = Core_k \setminus (\{x_k(op(Promise)) \} \cup learn_k(f,C))$, then $Cf'_k \subset Cl_0(Pre)$. Let $Cf_k$ be the non timestamped literals. Then if $S \vDash Cf_k$ and since $S$ is a state of a fair path, then if $p_{S,k}$ is the suffix path from $S$  but truncated of length $k$,  $p_{S,k} \vDash  Core_k\setminus x_k(op(Promise)) $, $UC_k(f) \setminus Presence_k(f)$ , $AUX_k(f),Learn_k(f,C),Next_k \vDash \neg x_k(op(Promise))$
\item $Cf_k = \{e \in Pre | N(e_0 = e)\rightarrow N(e_1)\rightarrow ...\rightarrow N(e_k=\neg x_{op(Promise)})$, with $ N(e_i )\rightarrow N(e_{i+1}) \in T_{black} $ and $N(\neg x_{op(Promise)}) \in SCC  \}$
\end{enumerate}
It is then straightforward that if $S \vDash x_{Promise} \wedge_{k \in \mathbb{N}}Cf_k$ then there is a contradiction since $p_{S}$ will never realize the operand promise $x_{op(Promise)}$. Furthermore, $\wedge_{k \in \mathbb{N}}Cf_k$ is computed as the set of $Pre$ contained in the backward fixpoint over $T_{Black}$ computing ancestors of any $\neg x_{op(Promise)} $  for all states of the SCC. 
\end{enumerate} 
\end{lemma}
\begin{theorem}
The learned clauses and Limit conflict clauses\footnote{in case of propositional conflict} are fair valid.\\
{ \bf sketch of the proof}:
By chronological induction on the  learned clause and limit conflict clause per conflict. First, assume that conflict $C$ is the first, thus the $Learn(f,C) = \emptyset $ at lemma \ref{next-valid}. Thus $\neg Cf$ and $\neg Limit_C$ is fair valid. Assume now that $Learn(f,C)$ are valid. Thanks to lemma \ref{next-valid}, it follows that $\neg Cf$  and $\neg Limit(C)$ are fair valid. 
%Thus, since there is a finite number of conflicts in the algorithm, then any learned clause is fair valid.
\end{theorem}

\begin{theorem}
The algorithm  terminates, is correct and complete\\
({ \bf sketch of the proof}):
\textit{ As long as a state is not known to be bad or in a Bad SCC, then it is recorded\footnote{by a hash table for instance} to avoid infinite loop. As soon as it is sure that it is a bad state or in a bad SCC, then a clause which will never be  forgotten and standing for the bad state is learned. Thus, our algorithm is similar to a depth-first-search of SCC in a LTL tableau \cite{KestenMMP93}. However, as soon as there is a conflict, the algorithm  prunes part of the tableau which is sure to lead to a failing state/SCC by, sound learning and backtracking using implication  dependencies of conflict.}
\end{theorem}

\begin{theorem}(Extraction of coarse small unsatisfiable core)\\
If $f = \wedge_i f_i$ then $\wedge_i \{f_i |x_{f_i} \in  \mathcal{A} (C_{last})  \}$ is a coarse small unsatisfiable core.\\ ({ \bf sketch of the proof}):
If the algorithm terminates with `unsat', the last conflict $C_{last}$  is at level 0. This means that ancestor  nodes in $\mathcal{A} (C_{last})$ with no parent gets a level 0, ie.  they correspond to some  clause in $presence(f)$ : $x_{f_i}$ where $f = \wedge_i f_i$ or eventually to some learned clause
 of the form $\neg x_h$. But since  $ICl(f,C_{last}) \setminus Presence(f) $ are fair valid, then $\wedge_i \{f_i |x_{f_i} \in  \mathcal{A} (C_{last})  \}
$ is a coarse small unsatisfiable core.\end{theorem}

\section{Conclusion}

In order to  detect which compliance rules are conflicting, we have provided a   conflict-driven Tableau depth-first-search for LTL. We have shown how it can be used to extract a small unsatisfiable core. Our method is theoretically $EXPTIME$ and $EXPSPACE$, but although deciding a MU is in $P-SPACE$ no $P-SPACE$ method have been proposed to extract cores yet. Our method does not suffer from cumbersome timestamped variables, handling of incrementation, searching upper bound for UMC. Implementation is ongoing work. Three  enhancements of the method would be to study a QBF-encoding of our method and analyzes if the learning we propose is easy for QBF solvers to learn. Other ways could be to use symbolic DFS \cite{BiereCZ99} or alternating B\"uchi automata.   
 Detecting conflicts in rules is critical for human interactive contract management systems. Moreover, our method pinpoints temporal issues in any automatic tool which is sensitive to the consistency of many evolving heterogeneous policies such as regulatory laws, internal business rules, security or privacy. 
The extension of our method to deontic modality ~\cite{BroersenDDM04,FenechPS09} used in contracts appears straightforward, and we are also  focusing on this issue.

\bibliographystyle{eptcs}
\bibliography{biblio}

% that's all folks
\end{document}